\documentclass[12pt]{article}
\usepackage{amsmath,amsthm,amsfonts,amssymb,amscd}
\usepackage{graphics}
\usepackage[latin1]{inputenc}

\renewcommand{\thefootnote}

\begin{document}

\centerline{\large\bf A note on the electromagnetic irradiation in a holed spatial region }

\smallskip

\centerline{\large\bf - a space-time approach}

\vglue .4in

\centerline{\large\bf Luiz C.L. Botelho}

\vglue .3in

\centerline{Departamento de Matemática Aplicada}

\smallskip

\centerline{Instituto de Matemática, Universidade Federal Fluminense}

\smallskip

\centerline{Rua Mario Santos Braga, CEP 24220-140}

\smallskip

\centerline{Niterói, Rio de Janeiro, Brazil}

\smallskip

\centerline{e-mail: botelho.luiz@superig.com.br}

\vglue .5in

\noindent\textbf{Abstract:} We study the role of the homological topological property of a space-time with holes (a multiple connected manifold) on the formal solution of the electromagnetic irradiation problem taking place on theses ``holed" space-times.

On three apendixes additional to the bulk section 1, we present as well important studies on this irradiation problem on others mathematical frameworks.

\vglue .2in

\noindent{\large\bf 1.\,The space-time electromagnetic irradiation problem}

\bigskip

Let $M = R \times N$ be a space-time with spatial ``holes" inside (a multiple connected manifold) with $N$ denoting $R^3$ (or a compact manifold with 3D-holes inside. One important mathematical method (formal) problem is the electromagnetic irradiation problem on $M$. Let thus be given sources configurations with compact support on $N$ for the generation of this electromagnetic radiation, represented in Maxwell Theory by an enough continuously differentiable vector field $J_\mu(x)$ on $M$. The irradiation problem is to solve (at least formally) the electromagnetic potential wave equation
\begin{equation}
\left(\frac{\partial}{\partial t^2} - \Delta\right) A_\mu(x) \equiv \square\,A_\mu(x) = J_\mu(x) \tag{1}
\end{equation}
plus some standard boundary condition of Mathematical Physics (Dirichlet, Neuman or Robbin) and obviously initial datum conditions for compact orientable manifold spatial section $N$ or irradiation Somerfeld conditions at the asymptotic limit of $t \to \pm\infty$ on the electromagnetic potential for the case of non-compact $N$..

It is formally supposed that the $M$-homological properties are reflected on the propagation through a formal application of a space-time Hodge-Helmotz-De Rhan decomposition for the four vectors fields on eq(1) (see Appendix 2 for a non space-time proposal). Namelly:
\begin{equation}
A_\mu(x) = (\partial_\mu\varphi)(x) + \varepsilon^{\mu\nu\alpha\beta}\,\partial_\nu B_{\alpha\beta}(x) + A_\mu^{\rm top}(x) \tag{2}
\end{equation}
\begin{equation}
J_\mu(x) = (\partial_\mu j)(x) + \varepsilon^{\mu\nu\alpha\beta}\,\partial_\nu J_{\alpha\beta}(x) + J_\mu^{\rm top}(x) \tag{3}
\end{equation}
Here $(\varphi,j)$ are scalar fields on $M$, $(B_{\alpha\beta}, J_{\alpha\beta})$ rank-two anti-symmetric on $M$ and $(A_\mu^{\rm top},J_\mu^{\rm top})$ denote the associated ``topological harmonic" fields on $M$ ([1]). Note that $M$ has cylindrical ``holes", which turns $M$ a ``manifold" possesing a non trivial set of ``harmonic fields".

A natural gauge fixing on the gauge invariant wave equation eq(1) is choose to be the natural one $\partial_\mu\varphi \equiv 0$ on $M$. Note this step is necessary in order to make the D'Alembertian on eq(1) a formally inversible differential operator.

\footnote{From the mathematical point of view of the De Rhan-Hodge theorem, one should consider the Maxwell dynamical equations analitically continued to the Euclidean section of the analitical continuation of the manifold $M$ to the complex manifold $\mathbb{C}\times M$ ($i\mathbb{R}\times M$ is a pure euclidean manifold where the analitically continued Minkowskyan Fields $A_\mu$ and $J_\mu$, namely $A_\mu^\varepsilon = (i\,A_0,\vec{A})$, $J_\mu^\varepsilon = (i \rho \vec{J})$ and through the Euclidean forms $\sum\limits_{\mu=1}^4 A_\mu^\varepsilon\,dx_\mu$ and $\sum\limits_{\mu=1} J_\mu^\varepsilon\,dx^\mu$, posseses the standard decomposition associated to the De Rhan-Hodge theorem. Its reverse analitically continued forms coefficients are the formal objects on eq(2) and eq(3).
 
Precise conditions on the differential topological properties to allow on this operation are not studied in this work ([2],[4]).}

Another point worth to call attention is that the topological electromagnetic irradiation $A_\mu^{\rm top}(x)$ is not determined from eq(1) and it is thus fixed as a ``boundary condition" in our problem. So we can disregard its role in the irradiation problem by choosing from the beginning $A_\mu^{\rm top}(x) \equiv 0$.

Note that when the electromagnetic vector potential is in the form of eq(2), it possesses the additional gauge invariance
\begin{equation}
B_{\alpha\beta}(x) \to B_{\alpha\beta}(x) + (\partial_\alpha A_\beta - \partial_\beta A_\alpha)(x) \tag{4-a}
\end{equation}
together, of course, with the original $U(1)$
\begin{equation}
A_\mu(x) \to A_\mu(x) + \partial_\mu A(x) \tag{4-b}
\end{equation}
which has been already fixed by the condition 
$(\partial_\mu \varphi)(x) \equiv 0$.

So, we further impose the fixing gauge on the problem:
\begin{equation}
(\partial_\mu B^{\mu\nu})(x) \equiv 0 \tag{5}
\end{equation}

So in order to solve formally eq(1), we substitute the Hodge-Helmotz-De Rhan decomposition directly on the wave equation eq(1) with the above hypothesis and gauge fixing conditions already stated. As a result we get the following dynamical equation (with the constraint of divergenceless source $j=0$. Note that $(\partial_\mu J_\mu^{\rm top})(x) \equiv 0$ em $M$)
\begin{equation}
(\varepsilon^{\mu\nu\alpha\beta}\,\partial_\nu\,W_{\alpha\beta})(x) = J_\mu^{\rm top} - \square\,A_\mu^{\rm top} = J_\mu^{\rm top}(x). \tag{6}
\end{equation}
Here
\begin{equation}
W_{\alpha\beta}(x) \equiv (\square\,B_{\alpha\beta} - J_{\alpha\beta})(x) \tag{7}
\end{equation}

The solution of eq(7) is straightforward obtained through elementary tensor algebra. It reads as of as:
\begin{equation}
W_{\alpha\beta}(x) = \square^{-1}\big[\partial^\mu(^* J_{\mu\alpha\beta}^{\rm top})\big]. \tag{8}
\end{equation}
Here
\begin{equation}
^*\,J_{\mu\alpha\beta}^{\rm top} \equiv \varepsilon_{\mu\alpha\beta\rho}\, J^{\rho top}(x). \tag{9}
\end{equation}

In other words, formally we have the following (operational) expression for the anti-symmetric two rank irradiation tensor
\begin{equation}
B_{\alpha\beta}(x) = \square^{-1} (J_{\alpha\beta})(x) + \frac{1}{12}\,\square^{-2} \big[\partial^\mu(^* J_{\mu\alpha\beta}^{\rm top})\big]. \tag{9}
\end{equation}

Here the Inverse D'Alembertian $\square^{-1}$ and its operatorial square $\square^{-1}\circ \square^{-1} = \square^{-2}$ are exactly determined by the Potential Laplacian problem on $M$. Namelly, it is given by the (causal) integral kernel of the formal inverse of the D'Alembertian (see Appendix 1).

As a result one gets the (causal) formal result
\begin{equation}
A_\mu(x) = \varepsilon^{\mu\nu\alpha\beta}\,\partial_\nu\bigg\{\square^{-1}(J_{\alpha\beta}) + \frac{1}{12}\,\square^{-2} (\partial^\zeta(^* J_{\zeta\alpha\beta}^{\rm top}))\bigg\}. \tag{10}
\end{equation}

At this point it is worth call the reader attention that the ``topological harmonic" term of eq(10) is causal also since
\begin{align*}
&\qquad\qquad 
\frac{1}{12} \bigg\{\varepsilon_{\mu\nu\alpha\beta}\,\square^{-2}\,\partial^\nu\,
\partial_\zeta (\varepsilon^{\zeta\alpha\beta\sigma}\,J_\sigma^{\rm top})\bigg\}\\
&= \frac{1}{12}\,\square^{-2}
\big[\overbrace{\big(\eta^{\mu\zeta}\,\eta^{\nu
\sigma} + \eta^{\mu\sigma}\,\eta^{\nu\zeta}\big)\partial_\nu\partial_\zeta}^{\equiv \square^{+1}}
\big] J_\sigma^{\rm top}
\sim \frac{1}{12}\, \square^{-1}\, J_\mu^{\rm top} \tag{11}
\end{align*}

As an amazing result of our note one can see that the ``topological harmonic" component of the source also propagates. In other words: ``topology" also carries energy and has also a dynamical content as a conclusion of our inquiry.

The case of anisotropic irradiation problem in the path integral formulation is on progress ([4], [5]).

\vglue .3in

\centerline{\large\bf Appendix 1}

\bigskip

\centerline{\bf The Cauchy problem: Spectral resolution}

\vglue .2in

The explicit solution of the Cauchy problem for the Wave Equation on a 3D
spatial compact orientable manifold with a given Riemannian metric $g$ and appropriated initial conditions (Dirichlet conditions)
\begin{align*}
\frac{\partial^2 U(t,x)}{\partial t^2} &= (\Delta_g U)(t,x)\\
U(x) &= f(x) \in C^3(N)\\
U_t(0,x) &= h(x) \in C^2(N)\\
U(t,x)\big\vert_{\partial N} &= 0
\end{align*}
is given explicitly by
\begin{align*}
U(t,x) &= \square^{-1} \big((x,t),(x',0)\big)\binom{f(x')}{g(x')}\\
&= \sum_{n=1}^\infty \bigg\{f_n(\cos(\sqrt{\mu_n}t)O_n(x) + h_n \left(\frac{\sin(\sqrt{\mu_n}t)}{\sqrt{\mu_n}t}\right)O_n(x)\bigg\} \tag{13}
\end{align*}
\begin{align*}
-\Delta_g\,O_n(x) &= \mu_n O_n(x)\\
O_n(x)\big\vert_{\partial N} &= 0 \tag{14}
\end{align*}
\begin{equation}
f(x) = \sum_{n=1}^\infty f_n\,O_n(x) \tag{15}
\end{equation}
\begin{equation}
h(x) = \sum_{n=1}^\infty h_n\,O_n(x) \tag{16}
\end{equation}
The full writing of the formal D'Alembertian for external imputs is now standard (The Durhamell principle) ([2]).

\vglue .3in

\centerline{\large\bf Appendix 2}

\bigskip

\centerline{\bf The non space-time approach}

\vglue .2in

In this case we apply the Helmotz theorem only for the electromagnetic vector potential since the fourth component $A_0(x,t) = \phi(x,t)$ is a scalar function on $M$. The current writes in similar notation $J_\mu = (\rho(x,t), \vec{I}(x,t))$.

We have thus the spatial analogous of eq(2)-eq(3):
\begin{equation}
\vec{A}(t,x) = r \circ t (\vec{H}(t,x)) \tag{17}
\end{equation}
\begin{equation}
\vec{J}(t,x) = r \circ t (\vec{J}(t,x)) + \vec{J}^{\rm top}(t,x) \tag{18}
\end{equation}

Proceeding by the same steps on the bulk of this note, one arrives at the integral representations
\begin{equation}
\phi(x,t) = \square^{-1} (\rho(x,t)) \tag{19}
\end{equation}
\begin{align*}
\vec{A}(x,t) &= \square^{-1} (r \circ t\, \vec{J})\\
&\,\,\, + \square^{-1} \bigg\{r \circ t \big[(-\Delta)^{-1}(r\circ t\, J^{\rm top})\big]\bigg\} \tag{20}
\end{align*}

As again, one may interpret physically the last term in eq(20) as the electromagnetic propagation due to the non trivial spatial topology of the ``holed" domain $N$.

\newpage

\centerline{\large\bf Appendix 3}

\bigskip

\centerline{\bf Electromagnetic irradiation problem on the approach of differential forms}

\vglue .2in

In this appendix, following closely the study of the formulation of Electrodynamics in term of forms ([3]), we present the study of this note in terms of differential forms.

The Maxwell equations on $\mathbb{R} \times \mathbb{R}_g^3$\,, where the spatial part $\mathbb{R}_g^3$ possesses a Riemannian metric $g$, is given by
\begin{equation}
d(^* F^2) = 4\pi\,S^3 \tag{21}
\end{equation}
\begin{equation}
d(F) = 0 \tag{22}
\end{equation}

Here the differential two-form associated to the electromagnetic strenght fields $(\vec{E}, \vec{B})$ reads as
\begin{equation}
F^2 \equiv \widehat{E}^1 \wedge dt + \widehat{B}^2 \tag{23}
\end{equation}
\begin{equation}
\widehat{E}^1 \equiv \vec{E}_1\,dx^1 + \vec{E}_2\,dx^2 + \vec{E}_3\,dx^3 \tag{24}
\end{equation}
\begin{equation}
I_{\vec{B}}({\rm vol}^3) \equiv \widehat{B}^2 \equiv B_{23}\,dx^2 \wedge dx^3 + B_{31}\,dx^3 \wedge dx^1 + B_{12}\,dx' \wedge dx^2 \tag{25}
\end{equation}
and the source differential three form
\begin{align*}
S^3 &\equiv \rho\,dx^1 \wedge dx^2 \wedge dx^3\\
&\quad -(j_1\,dy \wedge dz + j_2\,dz \wedge dx + j_3\,dx \wedge dy) \wedge dt \tag{26}
\end{align*}

By a direct application of the Hodge-De Rhan decomposition theorem to the above written forms and by taking into account the change conservation $(dS^3\equiv0)$ on the first Maxwell equation, one gets
\begin{equation}
S^3 = d(s^2) + h^{3,{\rm top}} \tag{27}
\end{equation}
\begin{equation}
^* F^2 = d(f^1) + d^*(f^2) + f^{2,{\rm top}} \tag{28}
\end{equation}
\begin{equation}
d^*\,f^3 = 4\pi d(s^2) + 4\pi h^{3,{\rm top}} \tag{29}
\end{equation}

By calling $\overline{\omega}_{\rm top}^2 = d^*f^3 - 4\pi s^2$, one can (in principle) determine its exact functional (coefficients) form by the duality relationship for differential forms that for any $\sigma^3 \in \Lambda^3(R\times R_g^3)$
\begin{equation}
\langle 4\pi h^{3,{\rm top}},\sigma^3\rangle_g = \langle d\omega^2, \sigma^3\rangle_g = \langle \omega^2, d^* \sigma^3\rangle_g \tag{30}
\end{equation}

By taking into account now that $df^3\equiv 0$, and the gauge fixing $f^1\equiv 0$ and $f^{2,{\rm top}} \equiv 0$ one finally has that

\begin{equation}
\Delta f^3 = d(\overline{\omega}^2_{\rm top} + 4\pi s^2) \tag{31}
\end{equation}
where $\Delta$ denotes the Laplacian acting on differential three forms on $\mathbb{R} \times \mathbb{R}_g^3$.

As a result, the irradiation field form $f^3$ has a component due to the harmonic piece of the form source $S^3$
\begin{equation}
f^3 = \Delta^{-1} (d(\overline{\omega}_{\rm top}^2)) + 4\pi\,\Delta^{-1}(s^2) \tag{32}
\end{equation}
\begin{equation}
^* F^2 = d^*(f^3) \tag{33}
\end{equation}

\vglue .1in

\noindent\textbf{\large Acknowledgments:} We are thankfull for a CNPq fellowship and to the colleague Professor W. Rodrigues for discussions on the Appendix 3.

\vglue .2in

\centerline{\large\bf References}

\medskip

\begin{itemize}
\item[{[1]}] Luiz C.L. Botelho - Lectures Notes in Applied Differential Equations of Mathematical Physics, World Scientific, New Jersey, 324 pp, (2008)
\item[{-}] Driver, B. and Frankel, T. - On the growth of waves in manifolds, J. Math. Anal. Appl. 178, pp 143-155, (1993)
\item[{[2]}] Duff, G., Spencer, D. - Harmonic tensors on Riemannian manifolds with boundary, Ann. Math. 56, pp 128-156, (1952)
\item[{[3]}] Frankel, T. - The Geometry of Physics - an introduction, Cambridge University Press - pp 654, (1997)
\item[{[4]}] Luiz C.L. Botelho - Random Operators and Stochastic Equations, V. 23, pp 53-67 (2015)
\item[{-}] Luiz C.L. Botelho - Random Operators and Stochastic Equations, V. 21, pp 271-290, (2013)
\item[{[5]}] Luiz C.L. Botelho - International Journal of Theoretical Physics, V. 49, pp 1396-1404, (2010).
\end{itemize}

\end{document}